\documentclass[letterpaper,twocolumn,10pt]{article}
\usepackage{usenix-2020-09}
\usepackage{svg}

\usepackage{tikz}
\usepackage{amsmath}
\usepackage{float}
\usepackage[normalem]{ulem}
\usepackage{filecontents}
\usepackage[numbers]{natbib}
\usepackage[font=small,labelfont=bf]{caption}
\usepackage{makecell}
\usepackage{algorithmicx}
\usepackage[noend]{algpseudocode}
\usepackage{multirow}
\usepackage{array}
\newcolumntype{C}[1]{>{\centering\let\newline\\\arraybackslash\hspace{0pt}}m{#1}}
\newcolumntype{L}[1]{>{\let\newline\\\arraybackslash\hspace{0pt}}m{#1}}
\usepackage[ruled,vlined]{algorithm2e}
\usepackage{epsfig}
\usepackage[table]{xcolor}
\usepackage{appendix}
\usepackage{titling}
\usepackage{fixltx2e}
\definecolor{mygreen}{rgb}{0, 0.788, 0.278}
\usepackage{verbatim}
\usepackage[export]{adjustbox}
\usepackage[font=small,labelfont=bf,textfont=bf]{subfig}
\usepackage{booktabs}
\usepackage{enumitem}
\usepackage{pifont}
\usepackage{balance}
\usepackage{wrapfig}
\usepackage{lipsum}
\usepackage{bm}
\usepackage{authblk}

\usepackage{amsmath}
\usepackage{amsthm}
\usepackage{xurl}
\usepackage{url}

\theoremstyle{definition}

\usepackage{xcolor} 

\usepackage[bottom,flushmargin,hang,multiple,para]{footmisc}
\renewcommand{\paragraph}{\textbf}

\newcommand{\sysname}{{ASAP}}
\newcommand{\hwcloud}{{\textit{Huawei Cloud}}}

\usepackage{tcolorbox}
\newtcolorbox{takeawaybox}{
    colback=gray!10!white,
    colframe=black!70,
    boxrule=0.8pt,
    arc=0pt, 
    boxsep=0pt,
    left=6pt, right=6pt, top=4pt, bottom=4pt
}

\begin{document}

\date{}

\pagestyle{plain}

\title{\Large ASAP: \underline{A} Di\underline{S}aggregated and \underline{A}synchronous Inference System for MoE \underline{P}refill  \vspace{-0pt}}

\author{Weiwei Chen, Shuang Chen\thanks{Corresponding author. Email: chenshuang0804@gmail.com}, Lele Li, Qiang Hu, Han Li, Xin Ye, Ming Yan, Zhibin Yu}
\affil{Huawei}
\maketitle

\begin{abstract}
Mixture-of-Experts (MoE) models have become the de facto standard for scaling large language models. To maintain computational efficiency, modern MoE serving systems typically employ a hybrid parallelism strategy, combining Data Parallelism (DP) for attention stages with Expert Parallelism (EP) for MoE stages. However, this design necessitates frequent global synchronization barriers between attention DP groups and experts. In online serving, significant variance in request arrival rates and sequence lengths inherently leads to \textit{DP imbalance}, causing severe synchronization stalls that degrade Time-to-First-Token (TTFT) and system throughput.

We present \sysname, an asynchronous inference system specifically designed to accelerate the prefill phase of MoE models. \sysname~ disaggregates the attention and MoE stages and implements a fully asynchronous execution pipeline. This is achieved through a suite of specialized asynchronous communication primitives and four coordinated optimizations across request scheduling and model execution, which collectively dismantle global synchronization barriers. We implement and evaluate \sysname~ on CloudMatrix384 super-nodes, demonstrating that it improves SLO-compliant prefill throughput by 90\% compared to state-of-the-art synchronous serving solutions.
\end{abstract}

\section{Introduction}

\paragraph{Background.} 
Large-scale Mixture-of-Experts (MoE) architectures have become the de facto paradigm for serving high-capacity models efficiently. By selectively activating a sparse subset of experts per token~\cite{gao2022parameterefficientmixtureofexpertsarchitecturepretrained,kimiteam2025kimik2openagentic,SurveyMOE,tang2025pangupromoemixture,SwitchTransformer,DeepSpeedMOE,GLaM}, MoEs decouple model capability from computational cost. For example, DeepSeek-V3~\cite{deepseekai2025deepseekv3technicalreport} achieves a total scale of 671B parameters while restricting active parameters to only 37B per token.

Since the vast majority of these parameters reside in expert weights, 
Expert Parallelism (EP) is commonly applied to the MoE stage, distributing experts across an array of accelerators. A wide EP configuration fundamentally reduces the memory footprint per device, directly alleviating the severe memory bottlenecks typical in LLM inference. Specifically, a DeepSeek-V3 prefill instance~\cite{deepseekai2025deepseekv3technicalreport} utilizes an EP size of 32, distributed across 32 GPUs over 4 compute nodes.

However, accommodating this broad EP scale introduces architectural friction for the attention stage, which must adopt a hybrid of Tensor Parallelism (TP) and Data Parallelism (DP). Due to the prohibitive inter-device communication overhead incurred by large TP, the TP degree is strictly bounded (typically $\le 8$). As a result, the system configures the DP size as $EP/TP$ to process concurrent request batches~\cite{MegatronLM, SequenceParallelism}. Consequently, the aforementioned DeepSeek-V3 instance enforces a TP size of 8 and a DP size of 4 during its attention phase.


\paragraph{Problem.} Despite the theoretical efficiency of these hybrid parallelism strategies, our profiling reveals severe resource underutilization and inflated inference latencies during the compute-intensive prefill phase of production MoE serving. This inefficiency stems from a fundamental execution mismatch: the \textbf{inevitable DP imbalance} coupled with the \textbf{rigid synchronization barriers} inherent in state-of-the-art inference systems. 

\begin{figure}[t]
\centering
\subfloat[Synchronous]{
\includegraphics[width=0.95\linewidth]{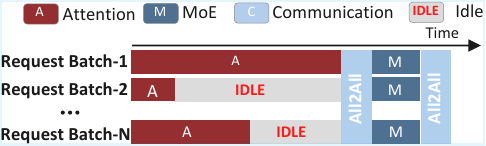}
\label{fig:example:sync}}\\
\subfloat[Asynchronous]{
\includegraphics[width=0.95\linewidth]{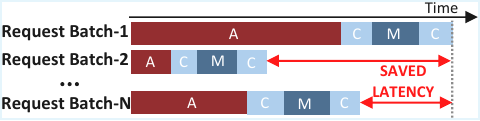}
\label{fig:example:async}}
\label{fig:sync-async-example}
\caption{Comparison between current synchronous and ideal asynchronous execution under heterogeneous request batches. Synchronous execution (Figure~\ref{fig:example:sync}) requires all request batches (launched on different attention DP groups) to synchronize before the MoE computation. An ideal asynchronous execution pipeline (Figure~\ref{fig:example:async}) eliminates the barrier, and allows faster request batches to progress at their own pace.}
\end{figure}

As depicted in Figure~\ref{fig:example:sync}, heterogeneous requests are independently batched and routed to distinct attention DP groups. However, because the subsequent MoE stage operates over a globally shared expert pool, these isolated DP groups are forced to synchronize at strict barriers before and after every MoE layer. This lockstep execution paradigm creates a severe \textbf{straggler effect}, where the entire system stalls until the slowest attention DP group or the most congested  expert completes its task. 

In production deployments, online inference clusters are  subjected to highly stochastic request arrivals, extreme sequence length skew, unpredictable prefix cache hit rates, etc~\cite{cai2026characterizing,xiang2025servegen}. Such  heterogeneity results in imbalanced workload (thus inconsistent latency) across attention DP groups, a phenomenon we term \textit{DP Imbalance}~\cite{liu2026revealing}.

\paragraph{Related Work.} Existing literature has extensively explored expert load balancing~\cite{deepseekai2025deepseekv3technicalreport,ma2025moegpsguidlinespredictionstrategy,cong2024predictionmoeneedsexpert}, yet the performance degradation caused by DP imbalance has been significantly underestimated in the context of online MoE serving. Because production LLM serving is characterized by highly stochastic arrivals and heavy-tailed sequence length distributions~\cite{xiang2025servegen, cai2026characterizing}, achieving perfect attention DP balancing is practically impossible. On another front, recent efforts in long-context acceleration~\cite{li2023lightseq, Bingyang2024LoongServe,ding2023longnet,huang2024advancing,MLSYS2025_2d04d975,MLSYS2025_7c180af0,FlexSP,child2019generating,katharopoulos2020transformers,hu2024memserve} successfully tackle super-long requests, but remain oblivious to the execution bubbles caused by heterogeneous short-to-medium requests (e.g., prompts with $<32$K tokens). Our characterization in Section~\ref{sec:back:attn} shows that DP imbalance does not necessarily come from super-long requests. Co-executing various short-to-medium requests in a synchronous inference system still stalls the pipeline and degrades overall throughput.


\paragraph{Motivation.} Figure~\ref{fig:example:async} illustrates an ideal, barrier-free, asynchronous execution paradigm. By eliminating global synchronization, heterogeneous requests are enabled to truly ``flow'' at their own pace without waiting for or delaying others. This approach effectively minimizes resource idling, which simultaneously reduces inference latency and improves overall throughput.

\paragraph{Challenges.} Transitioning from synchronous to asynchronous MoE serving presents  two fundamental architectural hurdles:
\begin{enumerate}[leftmargin=10pt,topsep=1pt,itemsep=-1pt]
\item \textbf{Decoupling Execution Dependencies with Asynchronous Communication Primitives:} Traditional systems tightly couple Attention and MoE stages on the same devices, enforcing a lock-step execution that precludes overlapping different computation stages. Transitioning to an Attention-MoE decoupled architecture requires not only physical resource disaggregation, but also a new class of asymmetric and asynchronous communication primitives. Current collective operations (e.g., Peer-to-Peer, All-to-All) are designed for synchronous blocking bulk transfers; the lack of efficient primitives that support fine-grained, non-blocking token routing is a primary reason why fully asynchronous MoE inference remains elusive.

\item \textbf{Maintaining Compute Efficiency under Fragmented and Out-of-Order Execution:} Asynchrony introduces "fragmented" execution; instead of one massive batch, MoE devices receive independent, smaller batches, risking a drop in arithmetic intensity. Furthermore, DP imbalance forces MoE devices to execute layers out-of-order, where the specific layer ID is only resolved at runtime. This dynamic execution pattern significantly increases kernel dispatch overhead from the host CPU, potentially introducing more execution bubbles than the gains obtained from asynchrony.

\end{enumerate}

\paragraph{Our Work.} We present \sysname, a high-performance inference system that realizes fully asynchronous MoE prefill. \sysname~ disaggregates attention and MoE stages onto separate devices, interconnected via asymmetric, asynchronous communication primitives to enable non-blocking data transfers. The key is to create a dedicated shared memory buffer on each device that is visible to all devices and acts as a ``superhub'' between senders and receivers. This is the foundation of \sysname's asynchronous execution pipeline. 

To fully leverage this asynchronous architecture and mitigate the efficiency risks of asynchrony, \sysname~ integrates four key runtime optimizations: (1) \textbf{Length-aware batching} to preserve MoE compute intensity; (2) \textbf{Dual-batch interleaving} to maximize attention device duty cycles; (3) \textbf{Communication-computation overlapping} via a triple-stream design to maximize hardware utilization; and (4) A layer-oblivious \textit{MoE Super Kernel} that enables \textbf{bubble-free kernel dispatching}, allowing out-of-order execution without host-side bottlenecks.

In summary, we make the following contributions:
\begin{enumerate}[leftmargin=10pt,topsep=1pt,itemsep=-1pt]
\item To the best of our knowledge, we are the first to build an asynchronous execution pipeline for online inference. 
\item We systematically analyze the prefill phase of MoE serving. We identify an inherent \textit{DP imbalance} problem in synchronous systems, causing severe resource idling and limited prefill performance.
\item We design \sysname, the first asynchronous MoE inference system that dismantles global synchronization barriers to enable  independent execution of request batches.
\item We implement and evaluate \sysname~on CloudMatrix384~\cite{zuo2025serving} supernode. Extensive evaluation results show that \sysname~improves SLO-compliant prefill throughput by 90\% compared to state-of-the-art synchronous baselines.
\end{enumerate}

\section{Background and Motivation}
\label{sec:background}

In this section, we provide the necessary background on MoE serving, and present a series of characterization studies to motivate the need for a barrier-free, asynchronous system to optimize MoE prefill performance. 

\begin{figure}[t]
  \centering
  \includegraphics[width=1\linewidth]{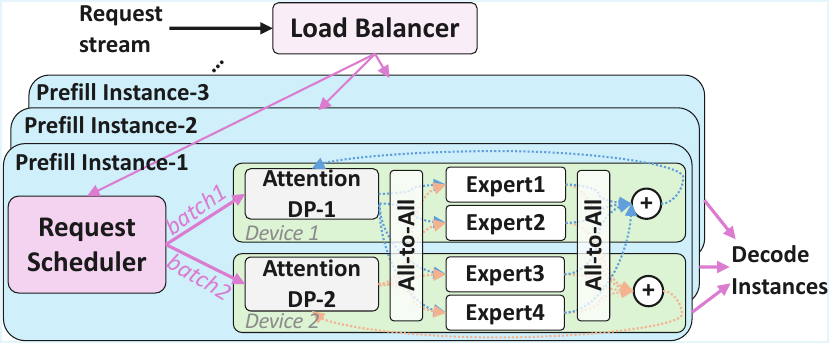}
  \caption{A typical MoE serving cluster consists of a number of serving instances. A user request is routed to a designated instance through a cluster-level load balancer. It is then aggregated into a request batch with other incoming requests, and dispatched to a specific attention DP group by the instance-level request scheduler. }
  \label{fig:bg:inference}
\end{figure}


\subsection{LLM Inference Preliminaries}
\paragraph{Prefill and Decode Phases.} In LLM serving, a user request first undergoes the prefill phase to generate the initial token, followed by the decode phase to generate subsequent tokens autoregressively~\cite{amali, sglang}. Inference latency is primarily characterized by Time-to-First-Token (TTFT) for prefill and Time-Per-Output-Token (TPOT) for decode. Since TTFT reflects how long the user has to wait before getting responses, its service-level objective (SLO) is usually within a few seconds~\cite{PastFutureScheduler,LazyBatching}. In contrast, SLO of TPOT is usually at tens of milliseconds, to keep up with the user's reading speed.

Given their divergent profiles---prefill being compute-bound and decode being memory-bound---it has become the standard practice to adopt prefill-decode disaggregation in production-grade inference frameworks~\cite{xiang2025aegaeon,deepserve,zhong2024distserve, Splitwise24}. 

\paragraph{MoE Architectures.} MoE models replace monolithic feed-forward networks (FFN) with a set of sparse, specialized sub-networks (experts). By activating only a fraction of parameters per token, MoE enables massive model scaling without a proportional increase in inference FLOPs. This architecture has rapidly become the de facto standard for state-of-the-art LLMs, as evidenced by the recent releases of DeepSeek-V3~\cite{deepseekai2025deepseekv3technicalreport}, Kimi-K2~\cite{kimiteam2025kimik2openagentic}, and Qwen3-MoE~\cite{yang2025qwen3technicalreport}.

\begin{figure}[t]
 \subfloat[Attention]{
   \includegraphics[width=0.5\linewidth,keepaspectratio]{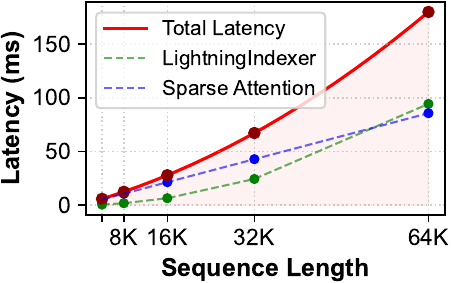}
 \label{fig:attn_latency}}
 \subfloat[MoE]{
   \includegraphics[width=0.48\linewidth,keepaspectratio]{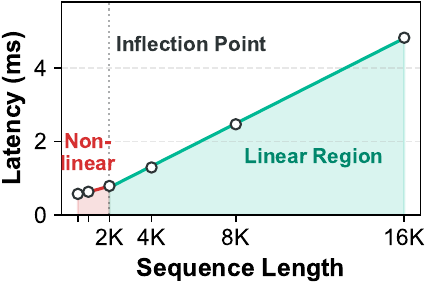}
 \label{fig:moe_latency}}
 \caption{Latency scaling with increasing sequence length.}
 \label{fig:back:cha}
 \end{figure}

 \begin{figure}[t]
  \centering 
  \begin{minipage}[t]{0.4\linewidth}      
    \includegraphics[width=0.96\linewidth,keepaspectratio]{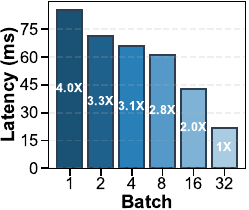}
    \caption{Impact of batch size on attention latency under a fixed total sequence length of 32k tokens.}
    \label{fig:attn_latency_bs}
  \end{minipage} 
  \hfill 
  \begin{minipage}[t]{0.57\linewidth}    
    \includegraphics[width=\linewidth,keepaspectratio]{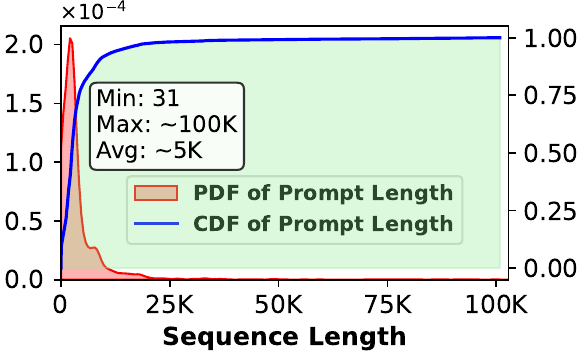}
    \caption{CDF and PDF of sequence lengths from a production dataset on \hwcloud, demonstrating extreme request variance.}
    \label{fig:CDF_of_token_length}
  \end{minipage}
\end{figure}

\paragraph{MoE Serving Infrastructure. }
Figure~\ref{fig:bg:inference} illustrates the typical architecture of an MoE serving cluster that is composed of a number of serving instances. Prefill-decode disaggregation is adopted, and we focus on illustrating the prefill phase (applicability to the decode phase is discussed in Section~\ref{sec:tech:discussion}). Upon the ingress of a user request, a cluster-level load balancer routes it to a designated prefill instance, typically following a round-robin policy. Within the assigned instance, the request is aggregated into a batch, and dispatched to a specific attention DP group by the instance-level request scheduler. After the first token is generated, the request along with its KV cache is dispatched to a designated decode instance to generate subsequent tokens.

To accommodate the massive parameter scale of MoE models, each serving instance typically employs wide Expert Parallelism (e.g., $EP=32$ in DeepSeek V3~\cite{deepseekai2025deepseekv3technicalreport}) to distribute experts across multiple accelerators, thereby alleviating the memory pressure per device ~\cite{deepseekai2025deepseekv3technicalreport, zhu2025megascale}. For the attention stage, Tensor Parallelism (TP) is usually applied to reduce attention latency. However, due to the large volume of communication, TP is generally limited to values of no more than $8$. To match the large size of the EP, the attention stage also adopts Data Parallelism (i.e., $DP=EP/TP$), as shown in Figure~\ref{fig:bg:inference}, . 

\begin{takeawaybox}
In this paper, we focus on optimizing performance of the prefill instances in MoE serving with a combination of attention DP and expert EP.
\end{takeawaybox}

\subsection{Characterization of MoE Inference}

To thoroughly analyze the runtime characteristics of the MoE prefill phase, we deploy DeepSeek-V3.2~\cite{liu2025deepseek}, a state-of-the-art open-sourced MoE model released in Dec 2025, on 32 Ascend NPUs~\cite{zuo2025serving}. Our setup follows the baseline configuration described in the DeepSeek technical report~\cite{deepseekai2025deepseekv3technicalreport}, utilizing a hybrid parallelism strategy of $EP=32$, $DP=8$, and $TP=4$. 
Comprehensive specifications of the model and hardware platform are detailed in Section~\ref{sec:eval:methodology}.

\subsubsection{Attention}
\label{sec:back:attn}

The attention stage is compute-intensive for the prefill phase. 
Figure~\ref{fig:attn_latency} plots attention latency as a function of sequence length $s$ for a batch size of one. It shows that the latency exhibits a quadratic correlation with $s$. Although DeepSeek-V3.2 incorporates DeepSeek Sparse Attention (DSA)~\cite{liu2025deepseek} where the sparse attention computation itself scales linearly, $O(s)$, the associated lightning indexer retains an $O(s^2)$ complexity. The combined effect of these operations dictates the overall quadratic scaling.

In production, request schedulers dynamically bundle multiple requests into a single batch to maximize hardware utilization. However, batching to a uniform total sequence length does not yield deterministic execution time. Figure~\ref{fig:attn_latency_bs} demonstrates the attention latency for a fixed total budget of 32k tokens across varying batch sizes. A batch size of 32 (1k tokens per request) compared to a batch size of 1 (32k tokens per request) reveals a stark latency disparity of up to 4.2$\times$. This significant variance arises because the attention complexity for a batch of $n$ requests is governed by the sum of the squares of individual sequence lengths, $O(\sum_{i=1}^{n}s_i^2)$~\cite{wang2025wlb}, rather than the square of their total sum.

\begin{takeawaybox}
Attention latency scales quadratically with individual request lengths. Consequently, the aggregate token count of a batch is a poor predictor of execution time.
\end{takeawaybox}

\subsubsection{MoE}
\label{sec:back:moe}
Conversely, the latency of the MoE stage is dictated primarily by the aggregate token count within a batch. 
Figure~\ref{fig:moe_latency} shows the scaling of MoE latency with $s$ under perfect expert load balancing. It reveals that MoE latency remains initially flat, transitioning to linear scaling only after surpassing an inflection point (approximately 2k tokens). This plateau occurs because the MoE stage is strictly memory-bound at low token count; the hardware must load massive expert weights from memory while performing relatively few computation operations. As sequence length grows, arithmetic intensity improves, pushing the system into a compute-bound regime. While the exact inflection point varies across hardware platforms and model configurations, this dual-regime scaling behavior—an initial memory-bound plateau followed by a linear compute-bound phase—is an intrinsic characteristic of MoE computation.

\begin{takeawaybox}
The MoE stage requires a minimum aggregate token count to saturate hardware compute units and transition from a memory-bound plateau to a high-efficiency linear regime.
\end{takeawaybox}

\subsection{DP Imbalance in Online Serving}

\label{sec:back:dp}
In production environments, inference clusters are subjected to highly stochastic request arrivals and extreme heterogeneity in sequence lengths, prefix cache hit rate, etc~\cite{cai2026characterizing,xiang2025servegen}. Figure~\ref{fig:CDF_of_token_length} visualizes the sequence length distribution from real-world \hwcloud~\footnote{A token service in production. The service and company names are kept anonymous for double blindness.}\ traces. While the mean prompt length sits at 5k tokens, the distribution exhibits a massive variance---ranging from a mere 31 tokens to a heavy tail extending up to 100k tokens. 
As established in Section~\ref{sec:back:attn}, excluding the impact of prefix cache, simply matching the total token count across attention DP groups is still fundamentally inadequate. Perfect load balancing would require that the sum of squared sequence lengths ($O(\sum s_i^2)$) precisely match, a constraint that is virtually impossible to satisfy dynamically. 


\begin{takeawaybox}
The combination of $O(s^2)$ attention complexity, heavy-tailed sequence length, and other factors (e.g., prefix cache) inherently causes latency variations across attention DP groups. Such DP imbalance leads to execution bubbles and resource idling in synchronous  systems.
\end{takeawaybox}

\section{\sysname\ Design}

In this section, we present \sysname, an asynchronous inference framework designed to accelerate the prefill phase of MoE models. By replacing the rigid barriers of conventional synchronous systems with a fully decoupled and asynchronous execution pipeline, \sysname~targets minimizing resource idling and maximizing prefill throughput under SLO constraints. To facilitate the subsequent discussion, Table~\ref{table:spec} lists all the symbols, definitions, and the representative configurations (detailed in Section~\ref{sec:eval:methodology}) used throughout this paper.

\subsection{Overview}
\begin{figure}[t]
  \centering
  \includegraphics[width=0.92\linewidth,keepaspectratio]{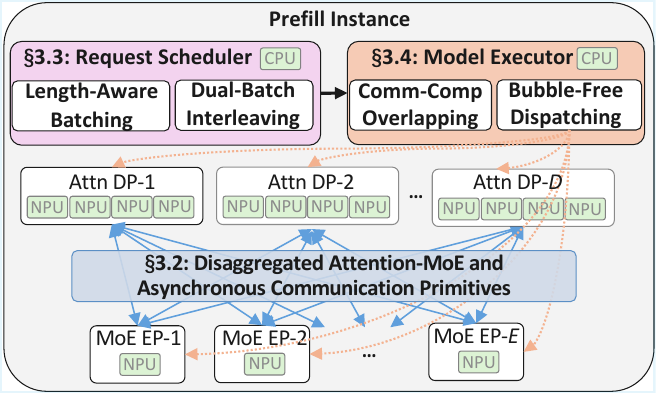}
  \caption{\sysname\ Overview. \vspace{-10pt}}
  \label{fig:system_overview}
\end{figure}


Figure~\ref{fig:system_overview} provides an overview of the \sysname\ architecture. \sysname~first adopts a disaggregated design by partitioning attention and MoE stages onto separate hardware devices. Specifically, the attention stage occupies $D\times T$ NPUs, while the MoE stage is deployed across $E$ NPUs
, resulting in a total hardware footprint of $D\times T + E$ devices per prefill instance.

To bridge these disaggregated components, we develop a suite of asynchronous communication primitives that support non-blocking 1-to-$N$/$N$-to-1 data transfers. This structural decoupling forms the foundation of our barrier-free inference pipeline, allowing the attention and MoE stages to progress independently without global synchronization.

Based upon this foundation, \sysname\ incorporates four key mechanisms within the request scheduler and model executor to maintain high hardware utilization:
\begin{enumerate}[leftmargin=10pt,topsep=1pt,itemsep=-1pt]
\item \textbf{Length-aware batching} to preserve MoE computational efficiency even under fragmented workloads;
\item \textbf{Dual-batch interleaving} to prevent attention devices idling during active MoE processing;
\item \textbf{Communication-computation overlapping} via a triple-stream design to maximize hardware resource utilization;
\item \textbf{Bubble-free kernel dispatching} to eliminate host-side dispatching overheads during out-of-order execution.
\end{enumerate}

\subsection{Asynchronous Communication Primitives}

\begin{figure*}[!t]
\centering
\subfloat[Asynchronous Dispatch]{
  \includegraphics[height=130pt,  width=0.47\textwidth, keepaspectratio]{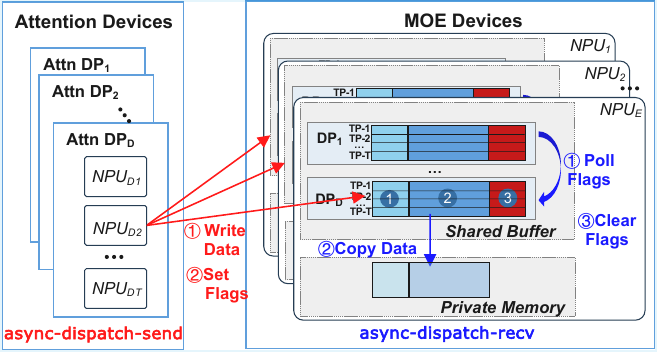}
\label{fig:async-dispatch}
}
\hspace{2em}
\subfloat[Asynchronous Combine]{
  \includegraphics[height=130pt,  width=0.45\textwidth, keepaspectratio]{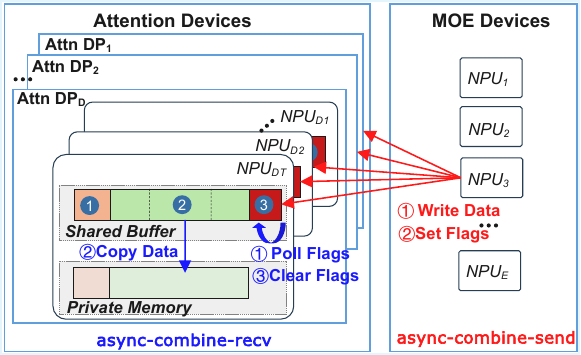}
\label{fig:async-combine}}
\label{fig:async-comm}
\caption{Asynchronous communication primitives and the shared buffer design. The buffer structure on each device is detailed in Table~\ref{table:buffersize}. Senders write data and set flags into the target device's shared buffer, returning to computation without synchronization. Receivers monitor these flags to retrieve and clear data as their own tasks progress.}
\end{figure*}


Conventional peer-to-peer and all-to-all collectives are inherently blocking, requiring explicit handshaking between senders and receivers. To eliminate this bottleneck, we implement a \textbf{distributed shared-memory abstraction}, where each device allocates a globally visible buffer as a decentralized communication hub. These buffers are \textbf{statically allocated} during initialization and persist throughout the framework's lifetime. Table~\ref{table:buffersize} details the buffer structure and memory footprint on each device. 

Under this model, senders write to the target buffer and immediately resume computation without waiting for receiver acknowledgment. Similarly, receivers independently retrieve data from these buffers upon completing local tasks. We leverage this abstraction and develop four specialized primitives, two for asynchronous dispatch and another two for asynchronous combine. These are designed to replace their synchronous counterparts, both of which are traditionally implemented as blocking all-to-all collectives. 

 \begin{table}[t]
\renewcommand{\arraystretch}{1.1}
\setlength{\tabcolsep}{5pt}
\footnotesize
\centering
\vspace{1pt}
\caption{Symbol Definition. Representative configurations are detailed in Section~\ref{sec:eval:methodology}.}
\label{table:spec}
\begin{tabular}{ccc}
\toprule
\multirow{2}{*}{\bf{Symbol}} & {\bf{Representative}} & \multirow{2}{*}{\bf{Definition}}  \\
 & {\bf{Configurations}} & \\
\hline
$D$ & 4 & Number of attention DP groups  \\
$T$ & 4 & TP in each attention DP group \\
$E$ & 16 & EP \\
$E_{total}$ & 256 & Number of experts in the model \\
$K$ & 8 & TopK in the model \\
$L$ & 61 & Number of layers in the model \\
$H$ & 7168 & Hidden size of the model \\
$Dsize$ & 16 bits & Data size \\
$S$ & 32k & Maximum batch sequence length \\
\bottomrule
\end{tabular}
\end{table}

\begin{table}[t]
\renewcommand{\arraystretch}{1.1}
\setlength{\tabcolsep}{2pt}
\footnotesize
\centering
\vspace{1pt}
\caption{Shared Buffer Structure. Example size is derived from the representative configurations specified in Table~\ref{table:spec}. }
\label{table:buffersize}
\begin{tabular}{c|c|c|c|L{1in}}
\toprule
& {\bf{ID}} & {\bf{Memory Size}} & {\bf{Example Size}} & {\bf{Description}}  \\
\hline
\multirow{3}{*}{\rotatebox{90}{Attention}} & \ding{172} & $K\times S/T$ & 64KB & Expert IDs \\
& \ding{173} & $H \times K \times S \times Dsize/T$ & 0.9GB & Expert results\\
& \ding{174} & $E/8$ & <1KB & E-bit bitmap flag \\
\hline
\multirow{3}{*}{\rotatebox{90}{MoE}}& \ding{172} & $D \times T \times E_{total}/E$ & <1KB & Token metadata \\
& \ding{173} & $D \times H \times K \times S\times Dsize$ & 14GB & Tokens (hidden states)\\
& \ding{174} & $D\times T/8$& <1KB & $D$ T-bit bitmap flags\\
\bottomrule
\end{tabular}
\end{table}

\subsubsection{Asynchronous Dispatch}

This occurs between an attention device and a set of MoE devices following each attention layer. Each MoE device allocates a shared buffer partitioned into $D$ regions (one per attention DP group), with each region further subdivided into $T$ rows (one per TP member within that DP group). 

As illustrated in Figure~\ref{fig:async-dispatch}, the shared buffer consists of three sub-regions:
\ding{172} token metadata stores token counts per expert ($E_{total}/E$ integers;
\ding{173} token payload stores the actual token hidden states; and
\ding{174} a $T$-bit bitmap indicating data readiness for each TP member.

\noindent \textbf{Execution Flow: } An attention device $NPU_{ij}$ invokes \\\texttt{async-dispatch-send} to write its local tokens into the $j$-th row of the $i$-th region in each target MoE device. To ensure data integrity, the sender employs a backpressure mechanism: if a target flag is already set, the write blocks until the receiver clears the flag. Once the transfer completes, the sender sets its corresponding bit in the bitmap (\ding{174}).

On the receiver side, the MoE device executes \\\texttt{async-dispatch-recv} by polling the bitmaps across all regions. When all $T$ flags for a specific $DP_i$ are set, the MoE device migrates the token metadata (\ding{172}) and hidden states (\ding{173}) to its private memory and clears the bitmap \ding{174} to acknowledge availability for subsequent transfers.

\subsubsection{Asynchronous Combine}
\label{sec:tech:ac}

This retrieves computed expert results from multiple MoE devices back to their originating attention DP group after each MoE layer. The attention device's shared buffer comprises: \ding{172} expert IDs for token-to-expert mapping; \ding{173} expert results partitioned into $E$ segments (one per MoE device); and \ding{174} an $E$-bit bitmap flag that  indicates result arrival from each MoE device.

\noindent \textbf{Execution Flow}: Upon completing the computation for $DP_i$, an MoE device invokes \texttt{async-combine-send} to write expert results into the shared buffer of the $T$ attention devices belonging to $DP_i$, and sets the corresponding completion bit. The attention device executes \texttt{async-combine-recv}, monitoring the bitmap flags. Once all activated expert results are received, the attention device migrates the payload to its private memory and clears the flags for the next iteration.

\subsection{Request Scheduler}

 While the asynchronous framework eliminates synchronization stalls, it introduces new challenges for maintaining high hardware utilization. Specifically, two primary issues arise. First, in a decoupled architecture, MoE devices process request batches sequentially rather than concurrently across all $D$ attention DP groups. This leads to reduced token density per MoE invocation, which may compromise computational efficiency. Second, resource disaggregation creates execution gaps; attention devices  become idle while their respective batches are in the MoE stage. 
 
 To address these challenges, we propose two scheduling optimizations as follows. 

\subsubsection{Length-Aware Batching}
\label{sec:tech:batching}

As characterized in Section~\ref{sec:back:moe}, the MoE stage requires a minimum token count (e.g., 2k tokens) to reach the compute-bound linear regime. Because the MoE stage processes batches from independent attention DP groups serially, maintaining a high per-batch token count is critical to saturate hardware utilization in an asynchronous environment. Our resource scheduler enforces a policy that aggregates requests into batches exceeding the inflection point in Figure~\ref{fig:moe_latency} (e.g., 2k tokens in our setup). Notably, since \sysname\ allows DP groups to progress independently, the batching algorithm is significantly simplified as it no longer needs to optimize for load balancing across attention DP groups.

\subsubsection{Dual-Batch Interleaving}
\begin{figure}[t]
  \centering
  \includegraphics[width=1\linewidth,keepaspectratio]{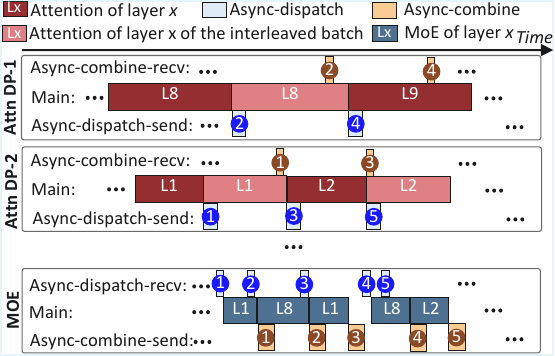}
  \caption{Dual-batch interleaving and the triple-stream design for communication-computation overlapping. For each attention DP group, two request batches are co-scheduled to interleave their layer-wise attention processing together. All devices employ a triple-stream design to further overlap asynchronous communication primitives with the main active computation kernels.}
  \label{fig:async_two_batch}
\end{figure}
\begin{figure*}[t]
 \begin{minipage}{0.6\textwidth} \centering
\subfloat[Per-layer MoE kernels.]{
  \includegraphics[width=0.38\linewidth]{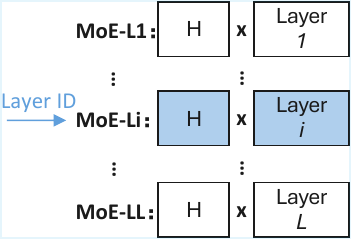}
\label{fig:layerless:before}} \hspace{15pt}
\subfloat[Unified MoE Super Kernel]{
  \includegraphics[width=0.47\linewidth]{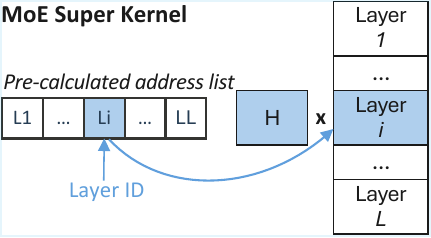}
\label{fig:layerless:after}}
\label{fig:layerless}
\caption{Comparison between standard per-layer MoE kernels and the proposed MoE Super Kernel. Block ``H'' denotes the tensor of hidden states, Block ``Layer\ i'' denotes the tensor of the expert weights of the $i^{th}$ layer.}
\end{minipage}
\hspace{10pt}
 \begin{minipage}{0.35\textwidth} \centering
\includegraphics[width=1\linewidth,valign=b]{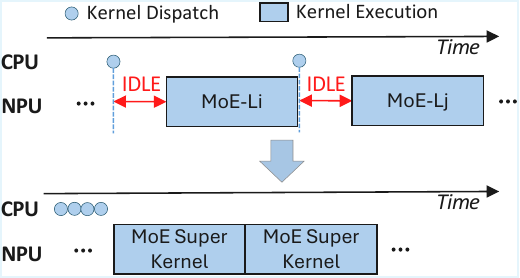}
\caption{The layer-oblivious design of the MoE Super Kernel allows ahead-of-time dispatching, eliminating device idling between kernels.}
\label{fig:layerless:host}
 \end{minipage}
\end{figure*}

To eliminate attention device idling during MoE execution, we porpose dual-batch interleaving. Rather than scheduling a single batch immediately upon reaching the inflection point, \sysname~waits for additional requests to form a dual-batch pair, and co-schedule the two batches to an idle DP group.

As illustrated in Figure~\ref{fig:async_two_batch}, the attention DP group interleaves the layer-wise processing of the two batches (the dark red and light red bars). When one batch is offloaded to the MoE stage, the attention devices immediately commence processing for the other batch of the same layer.

The efficiency of this interleaving depends on the relative latencies of the attention and MoE stages.
\begin{itemize}[leftmargin=10pt,topsep=1pt,itemsep=-1pt]
\item \textbf{Attention-limited cases}: If attention latency significantly exceeds MoE latency, each batch may experience queueing delays on attention devices. As shown in Figure~\ref{fig:back:cha}, for sequences exceeding 16k tokens, MoE latency accounts for less than 15\% of attention latency. Consequently, we disable interleaving for these long sequences, allowing them to exclusively occupy an attention DP group to minimize queuing from dual-batch interleaving.
\item \textbf{MoE-limited cases}: Conversely, if attention latency of the interleaved batch is shorter than MoE latency, then idling of attention devices will persist. However, with length-aware batching, attention execution typically surpasses several milliseconds, comfortably overlapping with the sub-millisecond to few-millisecond MoE latency.
\end{itemize}

\subsection{Model Executor}

The model executor serves as the core engine for inference execution. It orchestrates parallelism strategies, interfaces with the underlying hardware, and dispatches NPU kernels. Two primary runtime optimizations are proposed to minimize resource idling and maximize system throughput.

\subsubsection{Communication-Computation Overlapping}

Communication typically accounts for 10\% to 20\% of total execution latency~\cite{gond2025tokenweave}. To effectively mask communication overhead, \sysname~employs a \textbf{triple-stream concurrency model} on each device (Figure~\ref{fig:async_two_batch}). A dedicated compute stream executes the model's forward pass, while two independent communication streams handle asynchronous data transfers, ensuring that the hardware compute units remain saturated.

Specifically, the main compute stream of each attention device interleaves the layer-wise forward passes of the dual batches, while that of each MoE device processes MoE layers out-of-order. Each communication stream handles one of the four asynchronous primitives. By mapping these independent tasks to isolated hardware streams and allocate dedicated hardware resources to each stream, \sysname~approaches non-interfering execution (detailed in Section~\ref{sec:implementation}).


\subsubsection{Bubble-Free Kernel Dispatching}

As shown in Figure~\ref{fig:system_overview}, the MoE stage is a {\it shared} by multiple attention DP groups. Due to asynchronous processing, the MoE module executes layers in a nondeterministic, out-of-order manner. For instance, it may interleave the execution of Layer 8 from one DP group with Layer 1 from a newly scheduled batch in another DP group (Figure~\ref{fig:async_two_batch}). This layer execution order is highly nondeterministic, driven entirely by the independent progress of each attention DP group.

This dynamic execution order poses a significant challenge for kernel dispatching. Typically, host CPUs launch NPU kernels asynchronously in advance, effectively hiding the host-side dispatch overhead (e.g., argument preparation, memory allocation, tiling preparation, runtime API calls) behind active accelerator execution. However, because the exact layer ID to be executed next by the MoE stage is resolved dynamically at runtime, kernels cannot be pre-launched. This forces the CPU dispatch overhead---ranging from tens to hundreds of microseconds---onto the critical path (Figure~\ref{fig:layerless:host}). Given that individual MoE kernel execution spans merely sub-millisecond to a few milliseconds, this dispatch overhead constitutes a non-negligible performance bubble.

To eliminate these execution bubbles, we redesign the standard per-layer MoE kernel (Figure~\ref{fig:layerless:before}) into a unified layer-oblivious \textbf{MoE Super Kernel} (Figure~\ref{fig:layerless:after}). MoE kernel is typically a Grouped Matrix Multiplication (GMM), with hidden states being the left tensor, and expert weights of a specific layer being the right matrix. MoE Super Kernel introduces three key modifications:
\begin{itemize}[leftmargin=10pt,topsep=1pt,itemsep=-1pt]
    \item \textbf{Global Weight Access:} The kernel is granted pointer access to the expert weights across all $L$ layers. Because these weights are already statically resident in the MoE device's HBM, this modification incurs zero additional memory footprint.
    \item \textbf{Pre-calculated Address Indexing:} The memory offset for each layer's weight tensor is pre-computed and stored in an on-device address array for constant-time lookup.
    \item \textbf{Dynamic Resolution:} The layer ID is treated as a dynamic device-side argument rather than a host-side compilation constant. The kernel directly indexes the address array using the input layer ID to locate the correct target tensor before executing the GMM operation.
\end{itemize}

In this way, the MoE Super Kernel is oblivious of layer ID, allowing ahead-of-time kernel dispatching and eliminating device idling between MoE kernels (Figure~\ref{fig:layerless:host}). 

\subsection{Putting It All Together}

Upon arrival, user requests are aggregated via \textbf{length-aware batching} to ensure optimal MoE token density. Once a pair of batches is formed, they are co-scheduled to an idle attention DP group using \textbf{dual-batch interleaving}. The attention devices alternate their layer-wise processing, achieving seamless \textbf{communication-computation overlapping} by immediately computing the next request batch while invoking \texttt{async-dispatch-send} of the previous batch in separate communication streams. Concurrently, MoE devices process these decoupled workloads sequentially and out-of-order via the \textbf{MoE Super Kernel}, completely eliminating host-side stalls through \textbf{bubble-free kernel dispatching}. Finally, expert results are returned via \texttt{async-combine-send}, allowing the originating attention devices to eagerly advance to the next layer. Ultimately, this orchestrated pipeline fully dismantles global DP synchronization barriers, unlocking high-throughput, asynchronous MoE inference. 

\section{\sysname\ Implementation}
\label{sec:implementation}

\begin{figure}[t]
\centering
\includegraphics[width=0.78\linewidth]{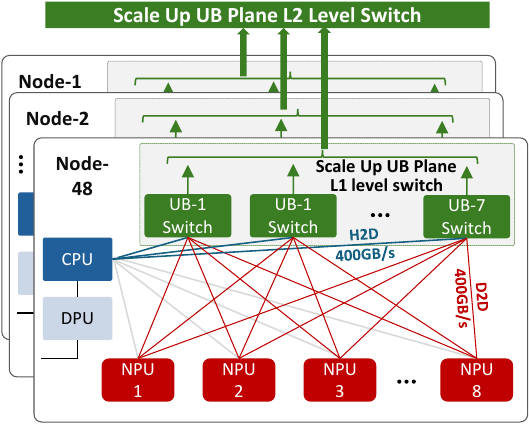}
\caption{Hardware architecture of the CloudMatrix384 supernode. The supernode comprises 48 nodes, each integrating 8 NPUs. All 384 NPUs are interconnected via a two-level hierarchical UB plane, providing 400~GB/s bandwidth between any NPU pair. } 
\label{fig:cloudeMatrix384} 
\end{figure}

\subsection{Hardware Platform}
\label{sec:impl:hardware}
\sysname~is implemented and deployed on the Huawei CloudMatrix384 supernode~\cite{zuo2025serving,deepserve} (Figure~\ref{fig:cloudeMatrix384}), which interconnects 384 Ascend 910 NPUs. Each NPU die has 24 cube cores, 48 vector cores and 64 GB HBM. The NPUs are linked via a Unified-Bus (UB)~\cite{ubmesh} mesh, providing a shared memory abstraction with a uniform bi-directional 400~GB/s bandwidth between any device pairs in the supernode. This architecture supports remote direct memory access with microsecond-level latency.

\subsection{Serving Framework and Deployment}
\label{sec:impl:framework}
\sysname~is integrated into a production-grade inference framework~\cite{deepserve} based on PyTorch 2.1 and CANN 8.3~\cite{CANN}. 
For our primary evaluation of DeepSeek-V3.2, we adopt a disaggregated Attention-MoE configuration with $D=4$, $T=4$, $E=16$, totaling 32 NPUs. This configuration maintains the same device count as the standard $DP8$-$TP4$-$EP32$ synchronous setup~\cite{deepseekai2025deepseekv3technicalreport}. Note that \sysname~supports other combinations of parallelism degrees, especially when $D\times T \neq E$ thanks to the disaggregated architecture. The optimal values of $D$, $T$, and $E$ can be derived from design space exploration~\cite{zhu2025megascale,deepserve}, which is orthogonal to our work.

Due to the HBM capacity constraints (which must host the KV cache on attention devices), we set the maximum sequence length per device to 8k. With $TP=4$, this allows individual request batches up to 32k tokens (i.e., $S=32k$); requests longer than $32k$ are routed to dedicated prefill instances utilizing Sequence Parallelism (SP). While our implementation and evaluation focus on these limits, \sysname's architecture is orthogonal to memory capacity and scales with available hardware resources.

\subsection{\sysname~Components}

\paragraph{Asynchronous Communication Primitives: } The four asynchronous communication primitives (\\{\tt async-dispatch-send},  {\tt async-dispatch-recv},\\ {\tt async-combine-send}, and {\tt async-combine-recv}) are implemented in AscendC~\cite{AcendC} with approximately 4,000 lines of code (LoC) in total. The shared memory buffer on each device is statically allocated using {\tt aclrtMalloc}~\cite{aclrtmalloc}. Reads and writes from one NPU to the remote memory buffer of another are called through {\tt DataCopy} or {\tt DataCopyPad} interfaces~\cite{datacopy} whose latency is at the granularity of microseconds.

\paragraph{MoE Super Kernel:} We extended the baseline MoE GMM kernel (about 3,000 LoC) to support layer-oblivious execution. By introducing a layer-ID input tensor and a pre-calculated weight address list (roughly 200 LoC), the MoE Super Kernel can resolve memory offsets dynamically at runtime.

\paragraph{Triple-Stream Design: } To mitigate resource contention between concurrent communication and computation, we partition hardware cores for our triple-stream design. Specifically, we allocate all 24 cube cores and 24 vector cores to the main compute stream. Each of the two communication streams is assigned 12 dedicated vector cores, which is sufficient to drive the non-compute-intensive data transfer logic. 

While this static partitioning isolates execution units, residual contention persists at the shared memory hierarchy, particularly within the L2 cache and HBM bandwidth~\cite{CANN}. Our empirical profiling reveals that on attention devices, even with core partitioning, concurrent communication-related memory traffic interferes with compute-intensive kernels, leading to a measurable elongation of attention latency. Consequently, we selectively deploy the triple-stream design only on MoE devices. 

\paragraph{Host-Side Logic:} We modified approximately 1,000 LoC in the request scheduler and model executor to implement length-aware batching, dual-batch interleaving, and bubble-free kernel dispatching.

\subsection{Discussion}
\label{sec:tech:discussion}
\paragraph{Portability Beyond Supernodes.} \sysname~does not strictly require supernode architectures. While UB-based load/store primitives provide ultra-low intra-node latency, the asynchronous paradigm is platform-agnostic. The logic remains robust over RDMA-based interconnects; even with RDMA latencies of hundreds of microseconds, the performance gains from dismantling global synchronization barriers (which can span tens of milliseconds) significantly outweigh the communication overhead.

\paragraph{Applicability to Decode Phase.} While \sysname~can be applied to the decode phase, the benefits are likely to be less pronounced for two reasons: (1) Complexity Scaling: Decode attention is $O(s)$ rather than $O(s^2)$, resulting in lower variance in attention latency and more manageable DP imbalance. (2) Arithmetic Intensity: The decode phase generates only one token per request. Accumulating enough tokens to saturate MoE efficiency (e.g., 2k tokens) would introduce queuing delays that may negatively affect decode SLO. Therefore, \sysname~primarily focuses on the prefill phase.

\section{Evaluation}
\label{sec:evaluation}

\begin{figure*}[t]
  \begin{minipage}[t]{0.39\textwidth} 
    \vspace{0pt} 
    \centering        
    \includegraphics[width=0.95\linewidth]{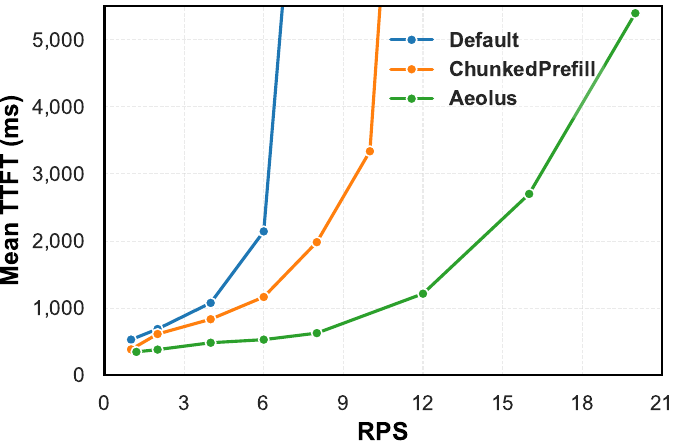}
    \caption{Mean TTFT with increasing request-per-second (RPS).} 
    \label{fig:overallPerformance}
  \end{minipage}%
  \hfill \hspace{1pt}
  \begin{minipage}[t]{0.25\textwidth} 
    \vspace{0pt}
    \centering    
    \includegraphics[width=1\linewidth]{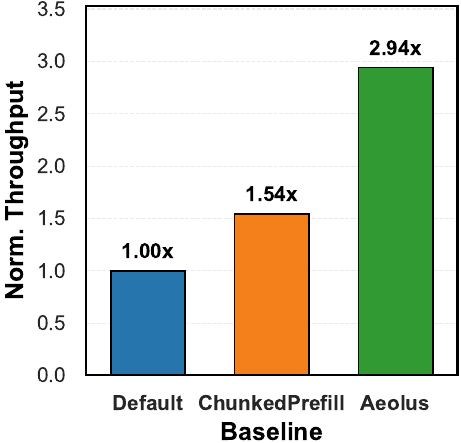}
    \caption{Normalized throughput under SLO.} 
    \label{fig:throughput_analysis}
  \end{minipage}
  \hfill 
  \begin{minipage}[t]{0.32\textwidth}
    \vspace{0pt} 
    \centering
    \includegraphics[width=0.97\linewidth]{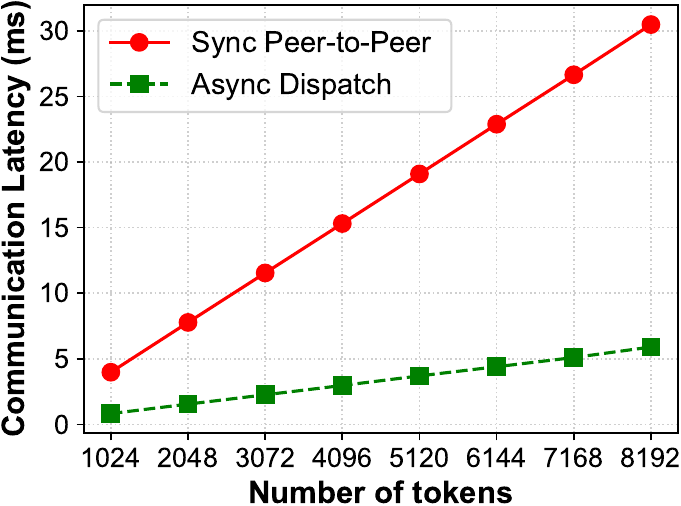}
    \caption{Communication latency with increasing token count.}
    \label{fig:dispatchVsP2P}
  \end{minipage}
\end{figure*}

\subsection{Methodology}
\label{sec:eval:methodology}
\paragraph{MoE Model.} We evaluate \sysname~using Deepseek-V3.2~\cite{liu2025deepseek}, a state-of-the-art open-sourced MoE model~\cite{liu2025deepseek}, released in December 2025. DeepSeek-V3.2 features a massive parameter scale of 671B, with 256 experts and 1 shared expert. It activates only 37B parameters (8 experts and 1 shared expert) per token. Deepseek-V3.2 incorporates DSA, a novel sparse attention mechanism.

\paragraph{Model Deployment.} Following the baseline deployment in Deepseek's technical report~\cite{deepseekai2025deepseekv3technicalreport}, one prefill instance of the synchronous baseline utilizes 32 NPU dies with $DP=8$, $TP=4$ and $EP=32$. To ensure a fair comparison, \sysname~employs a disaggregated $DP=4$, $TP=4$ and $EP=16$ configuration, maintaining the same device count as synchronous systems. All experiments are conducted on CloudMatrix384 supernodes (detailed in Section~\ref{sec:impl:hardware}).

\paragraph{Workloads.} Similar to prior work~\cite{Kwon2023PagedAttention,Bingyang2024LoongServe}, we evaluate various request rate and lengths. We simulate realistic serving scenarios with request inter-arrival times following a Poisson distribution~\cite{Poisson}. Request rate is quantified by request-per-second (RPS). Request lengths are sampled from a production trace dataset from \hwcloud's, shown in Figure~\ref{fig:CDF_of_token_length} and discussed in Section~\ref{sec:back:dp}.  Requests longer that 32k are excluded, as discussed in Section~\ref{sec:impl:framework}. We set the SLO for TTFT at 5 seconds.  Each run lasts for five minutes. 

\paragraph{Metrics.} All the systems decouple prefill and decode phases onto different devices, and we evaluate only the performance of the prefill instance. We evaluate: (1) \textbf{Mean TTFT} across varying request rates (RPS), and (2) \textbf{SLO-compliant Throughput}, defined as the maximum RPS sustained without violating the 5s SLO. 

\paragraph{Baselines.} We compare \sysname~against two synchronous state-of-
the-art inference systems: 
\begin{itemize}[leftmargin=10pt,topsep=1pt,itemsep=-1pt]
    \item \textbf{Default:} A standard scheduling and execution logic in the in-house inference framework~\cite{deepserve}, similar to vLLM~\cite{Kwon2023PagedAttention}. The resource scheduler aggregates incoming requests into batches of similar total token counts, to balance loads between DP groups.
    
    \item \textbf{ChunkedPrefill~\cite{agrawal2023sarathi}:} Long sequences are split into 8k chunks, which get processed sequentially. With $TP=4$, this means that the maximum sequence length on each attention device is reduced from 8k to 2k. This mitigates the sequence-length variance and DP imbalance problem.

\end{itemize}

\subsection{End-to-End Performance}

Figure~\ref{fig:overallPerformance} shows the mean TTFT of each inference system with increasing RPS, and Figure~\ref{fig:throughput_analysis} reports the normalized throughput under the 5s SLO. We observe that:
\begin{itemize}[leftmargin=10pt,topsep=1pt,itemsep=-1pt]
\item For each inference system, mean TTFT is flat at the beginning, and increases almost exponentially after an inflection point. \sysname\ significantly delays this inflection point. It consistently outperforms all the synchronous baselines across the entire range of request rates. For Default, load balancing using aggregate sequence length is proven to be ineffective, as discussed in Section~\ref{sec:back:attn}. For ChunkedPrefill, the reduced variance in sequence length indeed mitigates DP imbalance, thus outperforming Default. However, the system is still synchronous, and does not completely eliminate the DP imbalance issue.
\item At a low load ($RPS=1$) when requests rarely get queued in any of the inference systems, \sysname\ achieves mean TTFT of 350ms, representing a 34.3\% and 9.8\% reduction compared to Default (533ms) and ChunkedPrefill (388ms), respectively. As load increases, the benefits of asynchronous execution become more pronounced. At $RPS=4$, \sysname\ reduces TTFT by 54.9\% and 41.8\% over Default and ChunkedPrefill, respectively. At $RPS=8$, Default is unable to sustain the request rate, and \sysname~reduces mean TTFT by 68.3\% over ChunkedPrefill.
\item \sysname\ sustains an RPS of 20, whereas Default and ChunkedPrefill reach their limits at 6.8 and 10.5 RPS, respectively. This translates to \textbf{194\%} and \textbf{90\%} throughput improvement, demonstrating \sysname's superior ability to absorb heterogeneous workloads that traditionally trigger severe synchronization stalls.
\end{itemize}

\begin{figure}  
\centering
\includegraphics[width=1\linewidth]{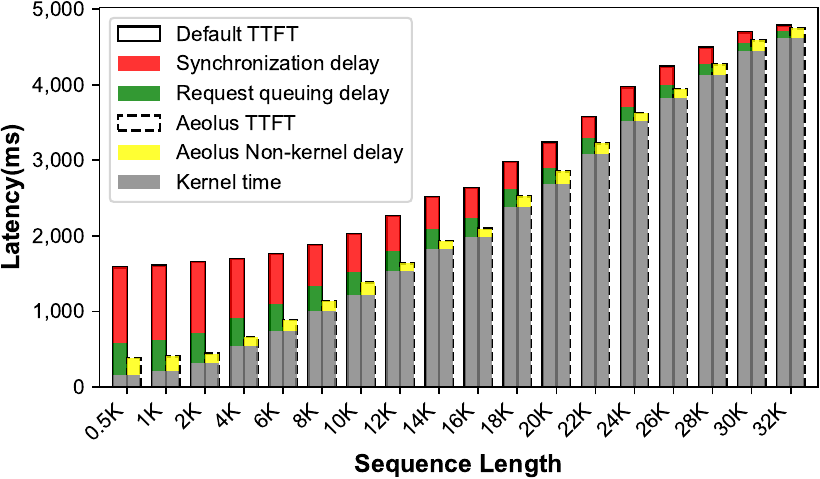}
\caption{Latency decomposition at QPS=4. The left and right bars at each sequence length show mean TTFT under Default and \sysname, respectively. For Default, TTFT is decomposed into kernel time, request queuing delay, and synchronization waiting delay. For \sysname, TTFT is decomposed into kernel and non-kernel delay.} 
\label{fig:gain_ayalysis} 
\end{figure}

\begin{figure*}[t] 
    \begin{minipage}[t]{0.305\textwidth}
        \vspace{0pt}
        \centering
        \includegraphics[width=1\linewidth]{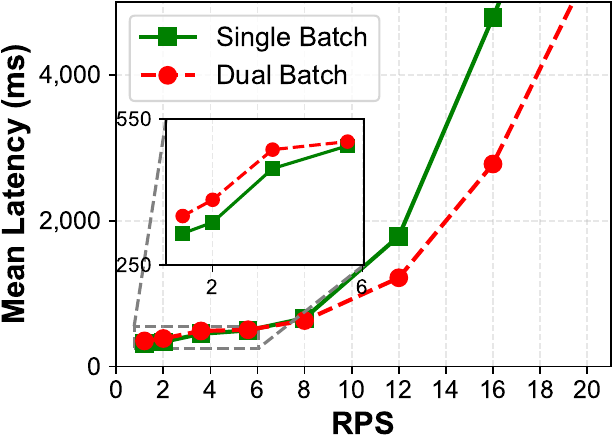}
        \caption{Comparison of mean TTFT with/without dual-batch interleaving.}
        \label{fig:CompareOneBatch}
    \end{minipage}
    \hfill
    \begin{minipage}[t]{0.322\textwidth}
        \vspace{0pt}
        \centering
        \includegraphics[width=0.99\linewidth]{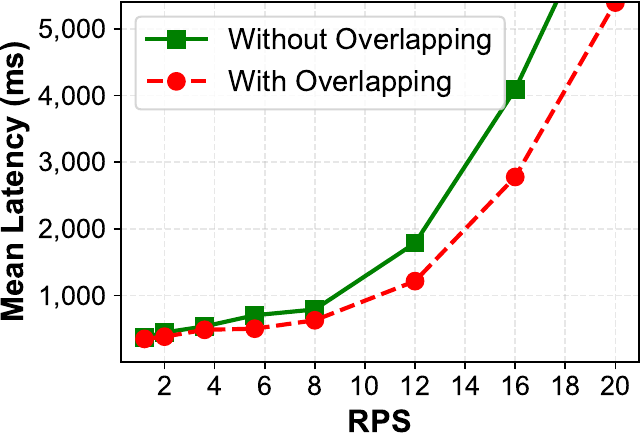}
        \caption{Comparison of mean TTFT with/without communication-computation overlapping.}
        \label{fig:ccoverlap}
    \end{minipage}
    \hfill
    \begin{minipage}[t]{0.3\textwidth}
        \vspace{0pt}
        \centering
        \includegraphics[width=1.02\linewidth]{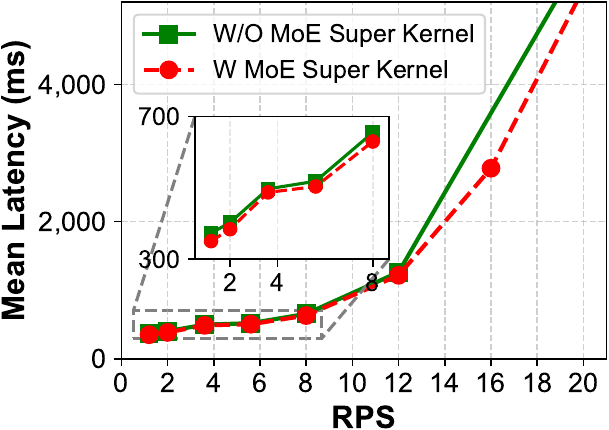}
        \caption{Comparison of mean TTFT with/without the bubble-free kernel dispatching brought by MoE Super Kernel.}
        \label{fig:b6Moepipelineavg}
    \end{minipage}    
\end{figure*}

\subsection{Latency Decomposition}

To understand the source of the performance gains, Figure~\ref{fig:gain_ayalysis} decomposes the TTFT of Default and \sysname~at RPS=4. for Default, TTFT is decomposed into kernel time (the actual computation latency), request queuing delay (waiting time after request arrival and before request processing), and synchronization waiting delay (due to DP imbalance). For \sysname, due to the small value, we do not further decompose non-kernel latency which mainly comes from queuing on attention devices due to dual-batch interleaving. Each bar represents a range of requests. For example, the first and second group of bars in Figure~\ref{fig:gain_ayalysis} represent mean latency of requests under 512 tokens, and requests between 512 and 1024 tokens, respectively. 

Latency increases with request sequence length due to increasing kernel time. Kernel time is the same under the two systems. However, in synchronous systems (Default), non-kernel overheads dominate for short sequences. For instance, for requests under 512 tokens, synchronization delay and request queuing delay account for 55\% and 30\% of the overall TTFT. This confirms the "straggler effect" where short requests are penalized the most by global barriers.

In contrast, \sysname\ reduces non-kernel delay up to \textbf{80\%} for the same sequence range. By allowing request batches to "flow" asynchronously, \sysname\ effectively converts idle synchronization time into productive computation intervals. As sequence length increases, the relative impact of synchronization diminishes, yet \sysname\ maintains a consistent edge by eliminating residual execution bubbles.


\subsection{Communication Efficiency}

The foundation of \sysname\ is its non-blocking, asymmetric and asynchronous communication. If not designing a new set of primitives, one has to use existing synchronous peer-to-peer (P2P) communication to realize asymmetric communication between attention and MoE devices. 

For instance, after attention computation, if the attention device needs to dispatch its tokens to $n$ MoE devices. Then, the P2P communication will be invoked $n$ times, Each time, the attention device (i.e., sender) has to handshake with each MoE device (i.e., receiver), and the data transfer will not be done until the receiver has received the data transfer. If the receiver is doing its own computation task, the sender will be blocked until the receiver finished the ongoing task and acknowledged the data transfer. Therefore, in p2p communication, the communication latency include (1) the handshake time which is a constant, (2) the data transfer time which is linear to data amount, (3) extra receiving delay which fluctuates depending on the status of each receiver, and (4) the total communication latency is $n$ times of the communication latency between the sender and each receiver. On the contrary, \sysname's \texttt{async-dispatch-send} is non-blocking, so it doesn't need to handshake with the receiver. It transfers data and immediately returns back future tasks. There is no extra receiving delay either because each receiver performs \texttt{asyn-dispatch-recv} whenever it's ready, without blocking the sender. Also, the sender can perform data transfer to $n$ devices at the same time in parallel.

Figure~\ref{fig:dispatchVsP2P} compares the communication latency of the synchronous P2P and the asynchronous {\tt async-dispatch}, with an increasing number of tokens. Both curves show a linear trend with token count. Sending $t$ tokens implies sending data size of $t*H*E_{total}/E$. For DeepSeek V3.2, this is 63MB per 1k tokens. We can see that latency of synchronous p2p is  \textbf{$4\times$} of {\tt async-dispatch} under 1k tokens, and \textbf{$5.8\times$} under 8k tokens. The difference increases with more data.

Note that the latency of our asynchronous communication primitive by leveraging supernode is significantly lower than that without support of the high-bandwidth peer-to-peer network in supernodes. Prior work~\cite{eaas} develops a set of asynchronous p2p primitives on GPUs leveraging the IBGDA Connection. The reported communication latency to send 512 tokens is over 0.5ms. However, the latency of our primitive is less than 0.1ms under the same token count.

\subsection{Ablation Studies}
We perform a series of ablation studies to show the effectiveness of each mechanism in \sysname.

\subsubsection{Effectiveness of Dual-Batch Interleaving}

Figure~\ref{fig:CompareOneBatch} evaluates the impact of dual-batch interleaving on mean TTFT across varying request rates. At low RPS, where request arrivals are sparse, the dual-batch mechanism introduces a marginal TTFT increase of approximately 6.5\%. This overhead stems from the difference in attention and MoE latency, as illustrated in Figure~\ref{fig:async_two_batch}: if the attention latency of one batch ($B_1$) exceeds the MoE latency of the co-scheduled batch ($B_2$), $B_2$ must wait for the attention devices to become available, introducing minor stalls.

However, as the RPS increases, the system without interleaving suffers from frequent attention-stage idling during MoE offloading, leading to rapid queue accumulation. By contrast, \sysname's dual-batch interleaving effectively reclaims these execution gaps, enabling the system to sustain high request rate under latency constraints. Consequently, it improves the SLO-compliant throughput by \textbf{14.3\%} (from 17.5 to 20 RPS). This result underscores that dual-batch interleaving is essential for maximizing attention device duty cycles and maximizing the overall system throughput.

\subsubsection{Effectiveness of Communication-Computation Overlapping}

The efficacy of our triple-stream design in masking communication overhead is analyzed in Figure~\ref{fig:ccoverlap}. In low-load scenarios, overlapping yields negligible TTFT reduction as the system operates with ample resource headroom. As the request rate intensifies, the overlapping becomes critical bottleneck, which increases the SLO-compliant throughput from 17.8 to 20, an \textbf{12.4\%} improvement.

\subsubsection{Effectiveness of Bubble-Free Kernel Dispatching}

Figure~\ref{fig:b6Moepipelineavg} quantifies the gains from the MoE Super Kernel that enables bubble-free kernel dispatching. At low RPS, the bubble-free kernel dispatching reduces TTFT by roughly 13ms. This is because the traditional kernel dispatching latency from the host CPU is 220 microseconds each layer measured on the CloudMatrix384 supernode. Since there are 61 layers in DeepSeek V3.2, this translates to $220*61\approx 13.4ms$ saving in TTFT. The latency saving also brings higher throughput. Compared to the systems without the MoE Super Kernel, our bubble-free kernel dispatching increases the SLO-compilant throughput by 6\%.
\section{Related Work}
\textbf{Long-context LLM Inference.} Extensive research has focused on accelerating attention computation for long sequences. 
FlashAttention~\cite{FlashAttention} and FlashDecoding~\cite{FlashDecoding} optimize memory I/O through tiling and recomputation, though they do not alter the quadratic complexity $O(s^2)$ to request sequence length $s$ in the prefill phase. Other approaches, such as sparse attention~\cite{child2019generating} and linear attention~\cite{katharopoulos2020transformers}, attempt to reduce this complexity. Our characterization (Section~\ref{sec:background}) and evaluation (Section~\ref{sec:evaluation}) have already applied sparse attention, and we demonstrate the necessity and benefits of \sysname's asynchronous execution. 

Sequence Parallelism (SP)~\cite{StripedAttention, li2023lightseq, RingAttention} partitions long sequences across devices to enable concurrent processing. 

LoongServe~\cite{Bingyang2024LoongServe} and Infinite-LLM~\cite{lin2024infinitellmefficientllmservice} further refine this by dynamically adjusting SP degrees or dedicating specialized clusters for long-context requests. Additionally, ChunkedPrefill~\cite{agrawal2023sarathi} and BucketServe~\cite{zheng2025bucketserve} mitigate request variance by decomposing long prompts or grouping similar-length requests. While these techniques are orthogonal and complementary to \sysname, they remain within the synchronous execution paradigm; they merely alleviate rather than eliminate the idle bubbles caused by intrinsic workload variance.
\\ \\ 
\noindent \textbf{Disaggregated LLM Serving Systems.} The computational disparity between different inference phases has led to the rise of disaggregated architectures. Prefill-decode disaggregation, exemplified by DistServe~\cite{zhong2024distserve}, Splitwise~\cite{Splitwise24}, and Aegaeon~\cite{xiang2025aegaeon}, separates the compute-bound prefill and memory-bound decode stages to minimize interference. Mooncake~\cite{mooncake} extends this by pooling CPU and DRAM resources for a distributed KV cache. However, these systems primarily focus on inter-phase isolation and do not address the intra-phase latency fluctuations driven by dynamic workloads.

Recently, Attention-MoE disaggregation has been proposed to enable independent scaling of attention and MoE modules, including frameworks such as MegaScale-Infer~\cite{zhu2025megascale} and Step-3~\cite{wang2025step}. However, they still rely on synchronous barriers across attention DP groups and do not tackle the DP imbalance dilemma. Furthermore, these frameworks are predominantly tailored for the decoding stage. In contrast, \sysname~ specifically targets the prefill stage, introducing an asynchronous execution pipeline to tackle resource idling due to DP imbalance.

\noindent \textbf{Asynchronous Inference Systems.} Expert-as-a-Service (EaaS)~\cite{eaas} designs a CPU-free asynchronous peer-to-peer communication library for MoE serving. However, it mainly targets scaling at the granularity of each expert, and does not tackle DP imbalance. The communication library includes only peer-to-peer primitives, and does not provide asymmetric 1-to-N/N-to-1 communication primitives. On the contrary, \sysname~adopts a set of efficient, asynchronous, and asymmetric communication primitives. Also, the communication latency of our primitives is much smaller than that reported in EaaS by taking advantage of supernodes.
\section{Conclusion}
Existing synchronous inference systems introduce costly barriers under DP imbalance in online MoE serving.
We present \sysname, the first asynchronous inference system to speed up the prefill phase of MoE models. 
Experiments show that \sysname~ improves throughput under SLO by 90\% over the state-of-the-art inference systems.

\bibliographystyle{plain}
\bibliography{references.bib}
\end{document}